\PassOptionsToPackage{table}{xcolor}
\documentclass[sigconf, 10pt, nonacm]{acmart}

\makeatletter
\@ifundefined{Bbbk}{}{}
\makeatother

\usepackage{tabularx}
\usepackage{makecell}
\usepackage{adjustbox} 
\usepackage{graphicx}
\usepackage{changepage}
\usepackage{subcaption}
\usepackage{tikz}
\usepackage[most]{tcolorbox}
\tcbuselibrary{listings}
\usepackage{csquotes}
\usepackage{amssymb}
\usepackage{amstext}
\usepackage{multirow}
\usepackage{booktabs}
\usepackage{pdfpages}
\usepackage{multicol}
\usepackage{placeins}
\usepackage{longtable}
\usepackage{enumitem}  
\usepackage{siunitx} 

\usepackage[capitalise]{cleveref}
\usepackage{framed}
\definecolor{shadecolor}{gray}{0.9}  
\newcommand{\fakepar}[1]{\vspace{1mm}\noindent\textbf{#1.}}
\newcommand{\fakepara}[1]{\vspace{1mm}\noindent \textit{#1}:}

\newcolumntype{L}{>{\raggedright\arraybackslash}X}   

\newcommand*\circled[1]{%
    \tikz[baseline=(char.base)]{%
        \node[shape=circle, fill, inner sep=2pt] (char) {\textcolor{white}{#1}};%
    }%
}

\newcommand{\circnumblack}[1]{%
  \tikz[baseline=(char.base)]\node[shape=circle,fill=black,inner sep=1pt,text=white] (char) {#1};%
}

\newcommand{\circnumred}[1]{%
  \tikz[baseline=(char.base)]\node[shape=circle,fill=red,inner sep=1pt,text=white] (char) {#1};%
}

\usepackage{xparse}

\tcbset{
  chatboxbase/.style={
    enhanced, breakable,
    arc=7pt, outer arc=7pt,
    left=2pt, right=2pt, top=3pt, bottom=0pt,
    width=\linewidth,
    fontupper=\footnotesize,                    
    fonttitle=\scriptsize\bfseries\color{white},
    shadow={0.4ex}{-0.4ex}{0pt}{fill=black!20},
    attach boxed title to top left={xshift=3mm, yshift=-2mm},
    boxed title style={
      boxrule=0pt,
      arc=6pt, outer arc=6pt,
      left=6pt, right=6pt, top=0pt, bottom=0pt
    }
  }
}

\NewDocumentEnvironment{chatlogexcerpt}{O{blue} O{Chat Log Excerpt} +b}{%
  \begin{tcolorbox}[
    chatboxbase,
    colback=#1!8,
    colframe=#1!60!black,
    title={#2},
    boxed title style={colback=#1!60!black, colframe=#1!60!black}
  ]%
  #3%
  \end{tcolorbox}%
}{}

\newcommand{\chatline}[2]{%
  \noindent\enquote{#1}
}

\def\BibTeX{{\rm B\kern-.05em{\sc i\kern-.025em b}\kern-.08em
    T\kern-.1667em\lower.7ex\hbox{E}\kern-.125emX}}
\begin{document}

\title{LLM-Enhanced Wearables for Comprehensible Health Guidance in LMICs}

\author{Mohammad Shaharyar Ahsan}
\email{25100169@lums.edu.pk}
\affiliation{%
  \institution{Lahore University of Management Sciences (LUMS)}
  \city{Lahore}
  \country{Pakistan}
}

\author{Areeba Shahzad Shaikh}
\email{25100046@lums.edu.pk}
\affiliation{%
  \institution{Lahore University of Management Sciences (LUMS)}
  \city{Lahore}
  \country{Pakistan}
}

\author{Maham Zahid}
\email{25100152@lums.edu.pk}
\affiliation{%
  \institution{Lahore University of Management Sciences (LUMS)}
  \city{Lahore}
  \country{Pakistan}
}

\author{Umer Irfan}
\email{27100363@lums.edu.pk}
\affiliation{%
  \institution{Lahore University of Management Sciences (LUMS)}
  \city{Lahore}
  \country{Pakistan}
}

\author{Maryam Mustafa}
\email{maryam\_mustafa@lums.edu.pk}
\affiliation{%
  \institution{Lahore University of Management Sciences (LUMS)}
  \city{Lahore}
  \country{Pakistan}
}

\author{Naveed Anwar Bhatti}
\email{naveed.bhatti@lums.edu.pk}
\affiliation{%
  \institution{Lahore University of Management Sciences (LUMS)}
  \city{Lahore}
  \country{Pakistan}
}

\author{Muhammad Hamad Alizai}
\email{hamad.alizai@lums.edu.pk}
\affiliation{%
  \institution{Lahore University of Management Sciences (LUMS)}
  \city{Lahore}
  \country{Pakistan}
}

\begin{abstract}
Personal health monitoring via IoT in LMICs is limited by affordability, low digital literacy, and weak health-data comprehension. We present \textit{Guardian Angel}, a low-cost, screenless wearable paired with a WhatsApp-based LLM agent that delivers plain-language, personalized insights. The LLM operates directly on raw, noisy sensor waveforms and is robust to the poor signal quality of low-cost hardware. On a benchmark dataset, a standard open-source algorithm produced valid outputs for only 70.29\% of segments, whereas Guardian Angel achieved 100\% availability (reported as coverage under field noise, distinct from accuracy), yielding a continuous and understandable physiological record. In a 96-hour study involving 20 participants (1,920 participant-hours), users demonstrated significant improvements in health data comprehension and mindfulness of vital signs. These results suggest a practical approach to enhancing health literacy and adoption in resource-constrained settings.
\end{abstract}

\maketitle



\section{Introduction}
\label{sec:intro}

Preventive health remains a critical, unmet challenge in Low- and Middle-Income Countries (LMICs)~\cite{haryanti2023comparative}. Restricted access to continuous, understandable health insights contributes significantly to the burden of preventable disease and delayed diagnoses in these regions. While personal health monitoring technologies have proliferated in resource-rich settings, their transformative potential in LMICs is stifled by a triad of socio-economic and technological hurdles: prohibitive costs of devices and data, low levels of digital and health literacy, and the substantial cognitive burden required to interpret health information~\cite{malik2020health, babalola2022health, perrins2024health}. Bridging this gap calls for personal health monitoring solutions that remain affordable and accessible while also being immediately comprehensible and actionable for users with diverse literacy levels in their current technological contexts.

\fakepar{Challenges} Four obstacles interact and reinforce one another, severely limiting both the adoption and the meaningful use of personal health monitoring technologies in LMICs.
\circled{1} \textit{The digital divide and interface complexity}: low digital literacy thwarts engagement with conventional mobile health platforms. This is often compounded by the complexity of these systems; feature rich graphical user interfaces impose steep learning curves, particularly for users with limited prior technology exposure or general literacy, thereby hampering navigation and the effective retrieval and interpretation of personal metrics~\cite{reid2024overcoming,kaphle2024systematic,guimaraes2023interface}; \circled{2} \textit{economic constraints}: with the high upfront cost of wearables inhibit individual adoption and large-scale deployment, especially where out-of-pocket healthcare spending dominates~\cite{huesch2024usage,swain2024analysis,wipo2024innovative,hpn2024public}; \circled{3} \textit{poor personalization and explanation}: as generic advice or raw data dashboards seldom clarify a metric's personal relevance, leaving users with low health literacy struggling to convert data into informed action~\cite{krebs2010meta,lee2025challenges,fadhil2024transforming,alqahtani2024recent}; and \circled{4} \textit{low cost sensor fidelity and algorithmic limits}: affordability pressures favour inexpensive, off-the-shelf sensors whose noisy signals overwhelm signal processing algorithms that are not designed to handle such high levels of motion artifact and noise. Motion artifacts and low signal-to-noise ratios yield missing or inaccurate readings, eroding user trust and utility. Consequently, existing personal health monitoring solutions, often designed without these deeply embedded constraints in mind, struggle for traction. Although prior work has attempted to tackle individual issues through simplified interfaces~\cite{guimaraes2023interface,choukou2022covid} or ultra-low-cost hardware~\cite{larnyoh2022using,yaakob2021cost}, \textit{a holistic solution that simultaneously addresses affordability, accessibility, personalization, and clear explanation in an integrated manner has remained elusive}.

\begin{figure*}[t] 
\centering
\includegraphics[width=\textwidth]{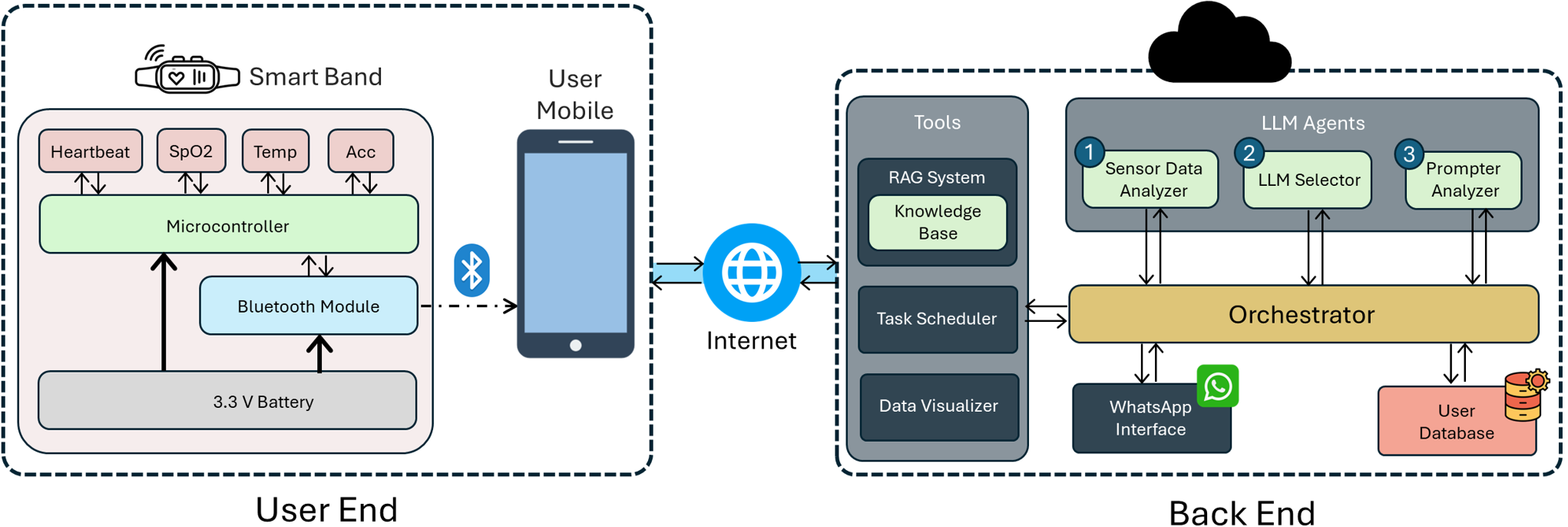}
\caption{System Architecture.}
\label{fig:system_architecture}
\end{figure*}

\fakepar{Approach} \label{subsec:approach}
Recent advancements in LLMs~\cite{alkhayat2024large,li2024systematic} present an opportunity to fundamentally re-imagine personal health monitoring accessibility, particularly by addressing the critical barriers of data comprehension and interface complexity. We introduce Guardian Angel, an end-to-end platform architected from the ground up to dismantle the interlocking barriers prevalent in LMICs. Guardian Angel (Figure~\ref{fig:system_architecture}) consists of a wearable health band and an LLM-driven WhatsApp chatbot working in concert ~\cite{lacroix2023using,conway2023improving,deniz2023quality}. The wearable captures vital signs at ultra-low cost, while the chatbot delivers personalized explanations and advice. We deliberately avoid a custom smartphone app; by leveraging WhatsApp, already ubiquitous in our target communities, we eliminate a major adoption barrier of learning new apps. Likewise, the wearable is screenless to minimize cost and complexity, offloading interpretation to the cloud where a tiered LLM pipeline analyzes data. Guardian Angel is a wellness aid, not a diagnostic tool; guidance is framed as conservative self-care cues.

\fakepar{Contributions}
Our work makes three key contributions to the design and evaluation of low-cost personal wellness monitoring in resource-constrained settings. 
\textbf{First}, we introduce a holistic architecture that fuses ultra-low-cost, screen-less hardware with a WhatsApp-based conversational agent driven by a tiered LLM pipeline, all tailored to the constraints and user needs of LMICs.  
\textbf{Second}, we focus on \emph{coverage under field noise} for low-rate commodity sensors: on a benchmark, a conventional quality-gated pipeline returned values for 70.29\% of windows, whereas our system produced outputs for 100\% of windows, reported as coverage, \emph{distinct from accuracy}, to maintain a continuous, understandable record.  
\textbf{Third}, in an exploratory in-the-wild deployment totalling 1920 participant-hours (20 participants, each continuously engaged for 96 h), we observe substantial improvements in health data comprehension and mindfulness of vital signs, even though pre-study surveys revealed low confidence in interpreting raw physiological data, a pattern common in LMICs~\cite{babalola2022health,lee2025challenges}.  

These contributions position Guardian Angel as a scalable, affordable pathway to democratize personal health monitoring and empower individuals to manage their well-being proactively, where such tools are most critically needed.


\section{System Design and Architecture}
\label{sec:design_arch}

Guardian Angel couples an ultra–low-cost edge with a resource-aware cloud to deliver usable health guidance in LMICs. A screenless wristband and minimalist phone app minimize BoM and power, while the cloud turns sensor streams into concise, actionable messages in familiar channels (e.g., WhatsApp). \textcolor{black}{Screen-less bands already exist, but most operate as closed-loop, proprietary stacks, preventing meaningful interaction or customization. Guardian Angel \emph{opens the loop} by decoupling sensing from interpretation and moving feedback to a programmable WhatsApp LLM agent, making interaction transparent and adaptable without changing hardware.} The contribution is the \emph{co-design} of a simplified edge with an adaptive backend that jointly addresses affordability, digital literacy, and comprehension barriers. We adopt a coverage-first design and position Guardian Angel as a wellness aid, not a diagnostic tool.

\subsection{Design Goals}
\label{sec:design_goals}
\begin{itemize}[leftmargin=*]
\item \textbf{Affordability and longevity:} prefer commodity parts and duty‑cycled sensing to extend lifetime on small batteries.
\item \textbf{Low cognitive load:} avoid device UI; deliver guidance in plain language in the user’s preferred channel and language.
\item \textbf{Robustness to noise and outages:} tolerate motion artifacts, intermittent connectivity, and missing samples without losing continuity.
\end{itemize}

\subsection{Wearable Band}
\label{sec:band_design}

\fakepar{Screenless, passive node}
We omit a display and on‑device menus, making the band a passive data node rather than a mini‑phone. This avoids UI complexity, lowers cost, and improves battery life. Interaction is offloaded to the mobile messaging interface. \textcolor{black}{Unlike prior screen-less bands that keep sensing and feedback locked together, we explicitly decouple sensing from explanation so the latter can evolve in the cloud without touching hardware.}

\fakepar{Sensor set and sampling}
With tight cost/power budgets, the band samples a small set of commodity sensors for wellness trends: PPG for pulse/SpO$_2$, skin temperature, and a 3‑axis accelerometer for activity/sleep. Sampling uses short fixed windows and simple on‑device filtering to suppress motion artifacts before BLE transfer. We intentionally avoid expensive sensors (e.g., ECG, GPS). Using commodity parts risks lower signal quality; the backend is explicitly engineered to tolerate such noise and maintain coverage.

\fakepar{BLE uplink}
BLE is the sole radio for pairing and uploading to the phone. It fits our duty‑cycled sampling, reduces peak power, and keeps the BoM low versus Wi‑Fi/cellular~\cite{montanari2017ble}. The firmware stages readings in a small FIFO and streams packets opportunistically to the background app when the phone is in range; otherwise, data is buffered locally until the next contact. Pairing uses authenticated BLE; payloads are sent over TLS via the app; device IDs are pseudonymous.

\subsection{Background Companion Application}
\label{sec:app_design}

The app bridges the passive band and the cloud service. On first launch, the user provides a phone number and a passcode. The app handles BLE pairing, time sync, and background pulls. It caches outgoing data, retries on network loss, and resumes interrupted uploads to survive low‑connectivity settings. The app stores only pseudonymous tuples (device ID, coarse timestamp, feature vectors). It exposes a minimal UI for consent, data export/delete, and alert preferences (e.g., daily summaries, escalation contacts). Users can pause uploads without unpairing.

\begin{figure*}[t]
  \centering

  \begin{minipage}{0.70\linewidth}
    \centering
    \includegraphics[width=\linewidth]{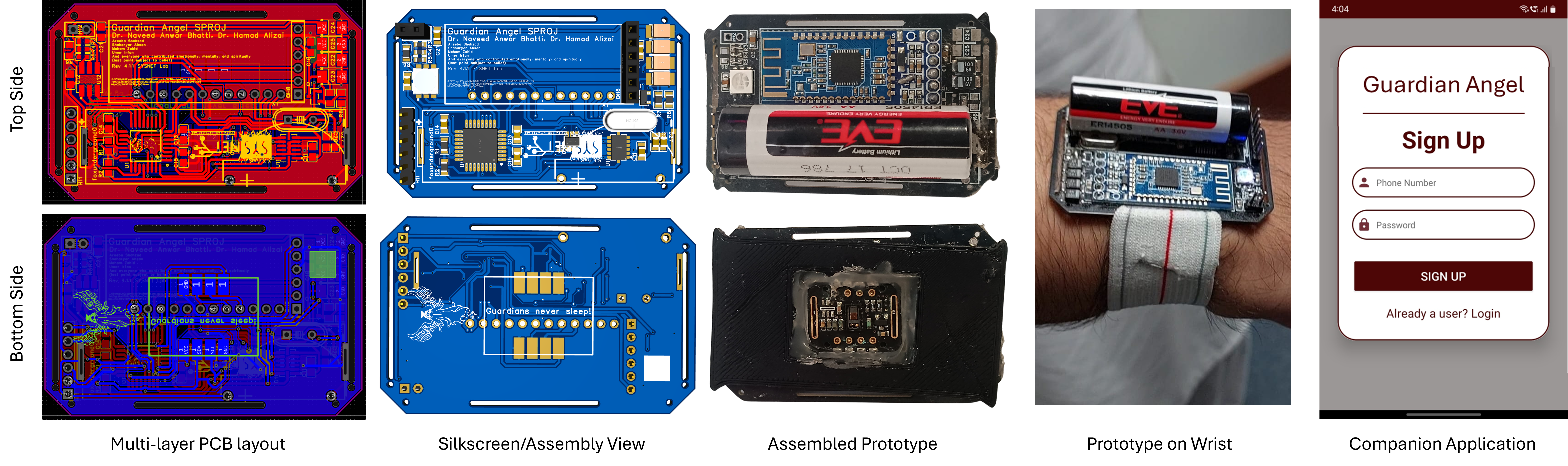}
    \caption{Guardian Angel: Design, assembly, and prototype in use.}
    \label{fig:schematic-grid}
  \end{minipage}
  \hfill
 \begin{minipage}{0.29\linewidth}
  \captionsetup{type=table,aboveskip=0pt,belowskip=2pt}
  \centering
  \footnotesize 
  \captionof{table}{BoM for the Guardian Angel.}
  \label{tab:sensor_pricing}

  \resizebox{0.75\linewidth}{!}{%
    \begin{tabular}{@{}l l S[table-format=2.2]@{}} 
      \toprule
      \textbf{Component} & \textbf{Role} & {\textbf{USD}} \\
      \midrule
      ADXL345         & Accelerometer   & 1.35 \\
      MAX30102        & Pulse oximeter  & 1.62 \\
      ATmega328P      & Microcontroller & 2.42 \\
      HM-10 BLE       & BLE module      & 2.85 \\
      ER14505 battery & Power (2.7 Ah)  & 1.78 \\
      Misc.\ passives & Caps/resistors  & 0.59 \\
      Custom PCB      & Circuit board   & 2.11 \\
      \midrule
      \textbf{Total}  &                 & {\bfseries 12.72} \\
      \bottomrule
    \end{tabular}%
  }
\end{minipage}
\end{figure*}

\subsection{Back-End Orchestration \& LLM Pipeline}
\label{sec:backend_design}

At the core of  Guardian Angel is an \emph{orchestrator} that supervises all services. When a sensor packet or a WhatsApp message arrives, the orchestrator spawns three LLM agents: \textit{Sensor Data Analyzer}, \textit{LLM Selector}, and \textit{Prompter Analyzer}, and hands them a shared context. It then provisions auxiliary tools requested by each agent: a retrieval augmented generation (RAG) interface backed by a medical literature index for clinical grounding, a cron-style scheduler for reminders, and an on-demand chart generator for trend visualization.

\fakepar{LLM-Driven Processing Stages}
\label{sec:llm_pipeline}
Although the orchestrator owns the control logic, the \emph{reasoning work} is divided into three sequential stages executed by the agents it launches:

\fakepara{Stage 1 – Sensor Data Analyzer}  
Runs a deterministic routine to derive HR, SpO\textsubscript{2}, temperature, and activity labels from each four-second sensor burst.  Performing this aggregation step shortens prompts and flags outliers early.

\fakepara{Stage 2 – LLM Selector}  
Inspects request complexity and urgency, then routes the task to the cheapest model able to answer it:
\textit{GPT-4o-mini} for look-ups, \textit{o3-mini} for multi-step trend reasoning, and \textit{o1} for safety critical interpretation.  
Section~\ref{sec:model-selection} quantifies the resulting $\sim$57\% cost reduction.

\fakepara{Stage 3 – Prompter Analyzer (Context\,\&\,RAG)}  
Assembles a structured prompt comprising (i) fresh clinical features, (ii) user profile + history, (iii) the last dialogue turns, and (iv) explicit output instructions.  
It then performs a RAG call to inject peer-reviewed medical guidance before passing the enriched prompt to the model chosen in Stage 2.  
The returned answer may also request a chart or schedule a reminder, in which case the orchestrator dynamically invokes the data visualizer or scheduler tool.



\section{Implementation}
\label{sec:implementation}
To support reproducibility, we release the wearable firmware/ hardware, Android app, backend (WhatsApp server), and evaluation scripts in an open-source repo on GitHub \cite{guardianangel_github}.

\subsection{Wearable Hardware and Firmware}
\label{sec:impl-hw}

The Guardian Angel wearable is built around an ATmega328P MCU \cite{atmega328p_datasheet}. The custom printed circuit board (PCB, Figure~\ref{fig:schematic-grid}) hosts three off-the-shelf sensors: an ADXL345 accelerometer \cite{adxl345_datasheet} for motion, a MAX30102 PPG module \cite{max30102_datasheet} for HR and SpO\textsubscript{2}, and a 10~k$\Omega$ NTC thermistor \cite{ntcm10k_datasheet} for both ambient and skin temperature. Communication is handled by an HM-10 BLE module \cite{hm10_datasheet}, and power is supplied by a 2.7~Ah ER14505 Li-SOCl\textsubscript{2} primary cell \cite{er14505_datasheet} chosen for longevity. Table~\ref{tab:sensor_pricing} details the specific components and associated prototype costs.

To balance sensing fidelity with battery life, we adopted specific sampling rates: 34.4~Hz for the accelerometer (sufficient to capture gait granularity and activity types \cite{khan2016optimising}), 31~Hz for the PPG sensor (adequate for reliable HR/SpO\textsubscript{2} waveform reconstruction \cite{bui4182587lossless}), and 1~Hz for the slower-changing skin temperature. These rates allow data collection in the MCU's 2~kB FIFO buffer, defining a 4-second acquisition window per cycle. We also implement a four-state loop to minimize power consumption: \textsc{Reset} $\rightarrow$ \textsc{Scan} (BLE advertising) $\rightarrow$ \textsc{Collect} (4-second sensor burst acquisition) $\rightarrow$ \textsc{Transmit}. After the \textsc{Collect} phase, the MCU powers down the sensors, activates the BLE module, and streams a 3.76~kB data packet at 9600 baud. A 3.5-second connection window plus a 1 ms inter-sample delay during transmission prevents BLE buffer overflows. The total transmission time per cycle is approximately 7.4 seconds. Between cycles, the radio enters a sleep state, targeting an average current budget below 500~$\mu$A, and extending battery life. Once transmission is complete or the connection window expires, the device returns to the \textsc{Scan} state.
Each BLE transmission contains timestamped sensor readings and metadata, structured as {\footnotesize\texttt{<ts, accel[128], ir[124], red[124], temp[4], CRC16>}} with authenticated pairing and pseudonymous device IDs.

\subsection{Android Companion App}
\label{sec:impl-app}

Android application utilizes the open-source FastBLE library \cite{fastble} for BLE management, handling device scanning, pairing, and encrypted communication. After initial pairing, where the band's unique BLE UUID is stored, a background service maintains the connection and manages the incoming sensor data queue. The app's sign-up screen (Figure~\ref{fig:schematic-grid}) is made to be minimal, only collecting the phone number and a password; remaining information is collected through conversational sign-up discussed in Section~\ref{sec:whatsapp}.

The raw ADC values in the data are then processed using: (i) a 3~Hz low-pass Butterworth filter smooths the temperature readings; (ii) a 0.5–2.5~Hz band-pass filter isolates the relevant frequencies in the PPG signals; and (iii) a 0.2 Hz high-pass filter removes the gravitational component from the acceleration data.

The processed sensor data, along with any anomaly flags, is serialized into a JSON object and transmitted securely via HTTPS POST request to the \texttt{/v1/sensors} endpoint on the backend server at 4-minute intervals, except in the case of anomalies, which are transmitted immediately. The application is designed for minimal resource consumption, typically idling below 1\% CPU usage and averaging around 6 MB of RAM. 

\subsection{Back-End LLM Processing Pipeline}
\label{sec:llm-pipeline}

\begin{figure*}[t]
    \centering
    \includegraphics[width=1\textwidth]{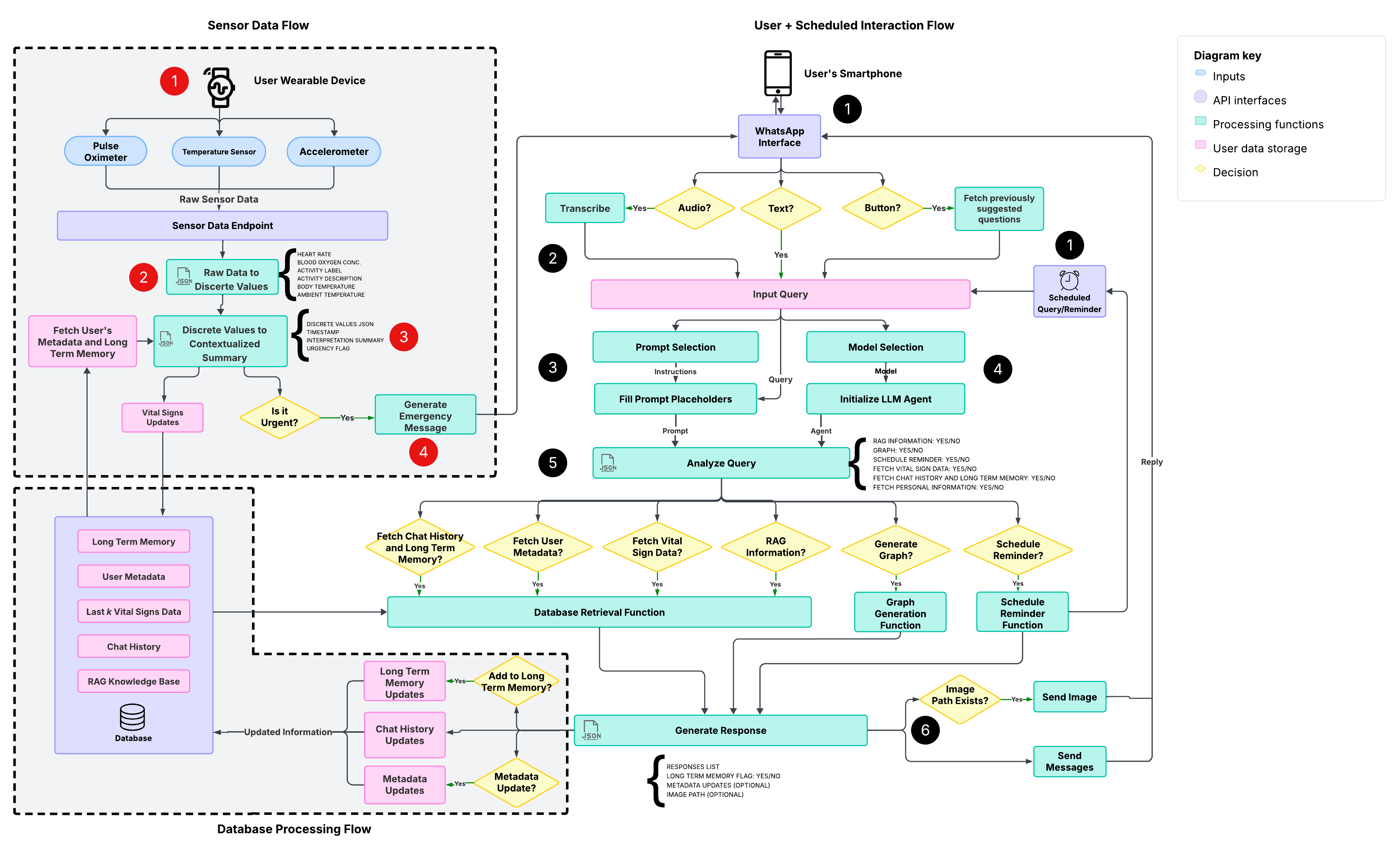} 
    \caption{System flowchart depicting dual data flows: Sensor Data Flow (left) and User/Scheduled Interaction Flow (right). }
    \label{fig:flow-diagram}
\end{figure*}

Figure~\ref{fig:flow-diagram} shows the end-to-end interaction flow, highlighting both the sensor-data path \circnumred{X} and the user-message path \circnumblack{X}.

\subsubsection{Adaptive Model Selection} 
\label{sec:model-selection}
High-performance LLMs (like o1) can incur significant inference costs, challenging scalability. To mitigate this, our system employs a tiered approach. Incoming user queries, received via the WhatsApp interface (right side \circnumblack{1}) and pre-processed if necessary (\circnumblack{2}), are first classified based on their inferred complexity. This classification is performed by a lightweight GPT-4o model acting as a router\footnote{All experiments used pre-GPT-5 models; GPT-5 released post-prep.}. Based on the router's classification, the system dynamically selects and invokes the most suitable LLM, balancing cost and capability (\circnumblack{4}):

\begin{itemize}[leftmargin=1em, labelsep=0.3em, itemsep=2pt]
    \item \textbf{Simple queries} (e.g., greetings, basic data requests) are routed to \textbf{GPT-4o-mini}.
    \item \textbf{Reasoning-based queries} (e.g., summarizing trends, requiring multi-step thought) are directed to \textbf{o3-mini}.
    \item \textbf{High-risk or complex queries} (e.g., interpreting potentially urgent anomalies, complex context) are escalated to \textbf{o1}.
\end{itemize}

This dynamic routing strategy ensures that computational resources are used efficiently, significantly reducing overall operational costs (as evaluated in Section~\ref{sec:cost_quality_analysis}).

\subsubsection{Adaptive Prompt Routing} 
A parallel prompt selection component (\circnumblack{3}) determines the requisite level of prompt specificity for the conversational agent. Simple utterances (e.g., greetings) trigger minimal prompt templates, whereas comprehensive health summary requests automatically trigger more detailed instructions to be embedded in the system prompt. This contextual routing is critical in both conserving computational resources and minimizing model hallucination by providing the LLM with only necessary and actionable contextual information.

\subsubsection{Structured Sensor-Data Analysis}
\label{sec:physio-interpreter}

Sensing streams arriving from the companion app (\circnumred{1}) traverse a two-stage pipeline:

\fakepar{Interpretation Module} (\circnumred{2}).
Low-cost personal health monitoring suffers from fragmented, intermittent data. Conventional pipelines drop noisy segments to preserve precision, which breaks continuity and undermines trust. We integrate an LLM as a pragmatic, instruction-following estimator over raw multi-modal waveforms. The model exploits correlations between accelerometer signatures and concurrent PPG artifacts to infer a plausible physiological state from segments a rule-based system would discard. The result is a coverage-first (continuity under field noise, distinct from accuracy), readable physiological stream that supports user trust and health literacy.

For each 4\,s burst, raw time series from IR/Red PPG (31\,Hz; 124 points per channel), the 3-axis accelerometer (34\,Hz; 408 points across \(a_x,a_y,a_z\)), and wrist/ambient temperature (1\,Hz; 4 points per sensor) are passed to GPT-4o-mini whose parameters are set as follows: {\footnotesize \texttt{temperature=1.3}, \texttt{top-p=0.8}.}

Unlike feature detectors that search for landmarks (e.g., systolic peaks in clean waveforms), the LLM acts as a sequence-to-values regressor. It processes the full multi-modal window, including accelerometer-linked noise in the PPG, and maps waveform shape to a plausible physiological estimate. Pre-training on diverse sequential data helps it see through noise that defeats deterministic algorithms. The prompt (Listing~\ref{box:llm-prompt}) directs the model to return `the HR and SpO\textsubscript{2} values you think it represents' and to `focus on outlying data', framing the task as informed estimation rather than exact feature extraction. The model outputs strict JSON: {\footnotesize\texttt{\{hr, spo2, activity, activity\_verbose, temp\_body, temp\_ambient\}}}, enabling a stream of numerical results validated in Section~\ref{sec:technical_evaluation}.

\begin{tcolorbox}[
  title    = Listing 3.3.3: LLM interpreter prompt,
  label    = box:llm-prompt,          
  colback  = gray!5,
  colframe = gray!60!black,
  fonttitle=\bfseries,
  fontupper   = \scriptsize,  
  left     = 2pt,
  right    = 2pt,
  top      = 2pt,
  bottom   = 2pt,
  listing  only,
  listing options = {
    basicstyle=\ttfamily,
    breaklines=true,
    columns=fullflexible,
    showstringspaces=false
  },
]
You will receive \texttt{ir} and \texttt{red} ppg data at 31~Hz, and
\texttt{a\_x}, \texttt{a\_y}, \texttt{a\_z} accelerometer data at 34~Hz,
as well as \texttt{body} and \texttt{ambient} temperature data at 1~Hz.

Return the \texttt{heart rate} and \texttt{SpO\textsubscript{2}} values you
think it represents, along with an \texttt{activity} label.
Also include one sentence suggesting what kind of activity the user might
be doing, as \texttt{"activity\_verbose"}.

Return the temperatures too.  Body temperature is taken at the wrist
(extremity, not core), so adjust if necessary.  If data are invalid,
return \texttt{"N/A"}.

Return a JSON object as the response.  Focus on outlying data.
\end{tcolorbox}

This end-to-end conversion builds on prior work \cite{xie2025physllmharnessinglargelanguage, time-series-linguistic-scaffolding, feli2025llmpoweredagentphysiologicaldata}. Although internal mechanisms are opaque, we hypothesize that broad sequential pre-training enables direct interpretation of raw physiological waveforms as an emergent capability. Accordingly, our evaluation focuses on empirical comparisons with conventional algorithms (Section~\ref{sec:technical_evaluation}).

    \fakepar{Reasoning and Urgency Module}(\circnumred{3}).  
    Extracted features are fused with longitudinal trends, demographics, and user-defined thresholds to generate (i) a concise health summary and (ii) an \texttt{URGENCY} flag (\textit{urgent}/\textit{not urgent}).  
    Urgent messages undergo an additional quality-control pass before delivery (\circnumred{4}); corresponding events are persisted as long-term memory alongside the vital-sign archive.

Together, these layers convert noisy sensor bursts and free-form chat into timely, personalised coaching, closing the loop from low-cost data acquisition to actionable insight.

\subsection{Dynamic Contextualization}

To anchor each reply in the user’s evolving context, the backend assembles a bespoke prompt immediately before every LLM call (\circnumblack{5}). The prompt embeds:

\begin{itemize}[leftmargin=1em, labelsep=0.3em, itemsep=2pt, topsep=0pt]
    \item \textit{Conversational History:} Ensuring context continuity across interactions.
    \item \textit{Long Term Memory:} Past events or medical alerts previously tagged as urgent to inform future interventions.
    \item \textit{Recent Health Metrics:} To set physiological context.
    \item \textit{Personal Profile Data:} User-specific information (name, age, BMI, medical background, demographic, phone number) to enable database lookup and contextual association.
    \item \textit{Temporal Context:} All queries and sensor records are furnished with their respective timestamps.
    \item \textit{Task Relevant Instructions:} To shape the nature and breadth of agent responses.
    \item \textit{Output Structure Specifications:} System prompts require all LLM output to conform to a structured JSON schema (e.g., \texttt{\footnotesize{PERSONAL, IMAGE, URGENCY, RESPONSES, QUESTIONS}}). Schema enforcement is managed via supported API configurations.
\end{itemize}

This dynamic scaffolding yields responses that are personalised, temporally grounded, and traceable to the user’s longitudinal record, crucial for safe, trustworthy health guidance.

\subsection{WhatsApp Interface Design} \label{sec:whatsapp}

End-users interact almost exclusively through WhatsApp (top-right \circnumblack{1}), which supports multilingual communication at scale. The backend handles five interaction facets:

\begin{itemize}[leftmargin=1em, labelsep=0.3em, itemsep=2pt, topsep=0pt]
    \item \textit{Input Processing:} The backend exposes webhooks for incoming messages of type \texttt{text}, \texttt{audio}, and \texttt{button}. Audio messages are processed by retrieving the media file, transcribing it via the OpenAI Whisper API, and converting the transcript to text.
    \item \textit{Conversational Sign-up:} When users sign up on the companion app, they are automatically sent a welcoming info message. It then prompts users to send personal information for contextual responses.
    \item \textit{Output Formatting:} Agent-generated output arrays (under \texttt{\footnotesize{RESPONSES}}) are broken into sequential WhatsApp messages; line breaks are explicitly handled for enhanced naturalness.
    \item \textit{Interactive Question Handling:} Follow-up questions, sometimes generated by the agent (i.e., \texttt{QUESTIONS}), are presented to the user with WhatsApp interactive buttons, affording high engagement and rapid iterative querying.
    \item \textit{Visual Data Delivery:} If the agent requests a visualization, the image file, generated through a server-side pipeline, is uploaded via WhatsApp's media API, accompanied by explanatory message content.
\end{itemize}

This design keeps all functionality within a familiar chat environment, eliminating the need for an app beyond one-time device sign-up.

\subsection{Agentic Abstractions and Tool Integration} \label{sec:agentic-tools}

Built on LangChain, our agent layer endows the LLM with concrete tool affordances that turn chat into action:

\begin{itemize}[leftmargin=1em, labelsep=0.3em, itemsep=2pt, topsep=0pt]
    \item \textit{Task Scheduling:} Scheduling is orchestrated through an external job service (cron-job.org) and an internal metadata registry. Callbacks trigger server-side actioning at scheduled times, supporting use cases such as medication or check-in reminders. Some tasks are scheduled by the system on sign-up, such as automated summaries and reminders if no data has been collected for a defined interval. 
    \item \textit{Data Visualization:} The agent can invoke a standardized plotting API, generating data visualizations.
    \item \textit{On-Demand Data Retrieval:} LLM agents are enabled to programmatically poll for the latest vital signs, user metadata, or timestamp context, facilitating grounded responses and reducing hallucination risk.
    \item \textit{RAG:} The architecture allows for expansion with a vector database (Pinecone) supporting retrieval-augmented workflows. Using multi-query retrieval, the system can synthesize knowledge both from general medical resources and user-uploaded documents, reinforcing the agent's factual grounding and personalization for wellness education.
\end{itemize}

These modular tools underpin robust planning, multimodal feedback, and fact-checked explanations, delivering a richer, more interactive wellness experience.

\section{Evaluation}

To assess the system’s feasibility, effectiveness, and potential impact, particularly in resource-constrained settings, our evaluation spans three areas: end-to-end system robustness, cost-efficiency and response quality, and user experience and impact. 
Guided by this scope, our evaluation focuses on where leverage is highest: the interpretation layer. In our target settings, success hinges on delivering continuous, comprehensible feedback at very low cost, not on shaving marginal analog tolerances. The interpretation layer (i) stays robust to commodity-part variance and wear/fit effects, (ii) turns noisy bursts into uninterrupted streams that sustain user trust, and (iii) transfers across devices and supply chains; fine-grained hardware micro-benchmarks, while valuable, are orthogonal to these deployment drivers and thus deliberately excluded from evaluation as they operate at a different level of analysis.

    
    

\subsection{Evaluating End-to-End System Robustness}
\label{sec:technical_evaluation}

We quantitatively compare an LLM with conventional signal processing for deriving HR, SpO\textsubscript{2}, and activity levels on a benchmark dataset. Key metrics are activity accuracy, mean absolute error (MAE) for HR and SpO\textsubscript{2}, and the fraction of segments with valid HR/SpO\textsubscript{2} estimates (data availability) under noisy PPG.

\subsubsection{Methodology}
\label{sec:eval_methodology}

We use the PhysioNet Pulse Transit Time PPG dataset \cite{ppg-dataset}, which provides three recordings per subject for 22 healthy subjects with time-aligned reference SpO\textsubscript{2}, HR, and IMU signals (iHealth Air pulse oximeter, OMRON monitor). The dataset employs the MAX30101 PPG sensor, matching \textit{Guardian Angel} hardware. Original signals (PPG at \SI{1000}{\hertz}, IMU at \SI{500}{\hertz}) are downsampled via linear interpolation to our device rates: \SI{31}{\hertz} for IR/Red PPG and \SI{34}{\hertz} for the tri-axial accelerometer. \textcolor{black}{This choice ensures sensor-family parity (MAX3010x) and trusted references, so accuracy numbers reflect algorithmic differences rather than mismatched optics or weak ground truth.} We target feasibility under commodity, low-rate operation; results should be read in this deployment context rather than as SOTA claims. We compare two processing pipelines:

\fakepar{Conventional Algorithmic Approach}
\begin{itemize}[leftmargin=1em, labelsep=0.3em, itemsep=2pt]
  \item \textit{HR/SpO\textsubscript{2} estimation:}
  A widely used open-source implementation \cite{hrcalc}, representative of typical signal-processing chains \cite{soriano2022design, singh2023measurement}. Steps include DC removal, filtering, IR peak detection, and AC/DC analysis on IR. Internal quality checks suppress outputs when signal criteria are not met.
  \item \textit{Activity classification:}
  An LSTM (50 units) trained on MotionSense \cite{Malekzadeh:2019:MSD:3302505.3310068} following prior work \cite{har-survey, zhao2017deepresidualbidirlstmhuman, odhiambo2022humanactivityrecognitiontime}. Input is tri-axial accelerometer (x, y, z) only; classes are \{sit, walk, run\}; early stopping is used.
\end{itemize}

\fakepar{LLM-Based Approach (Guardian Angel)}
Each 4\,s window is processed with the GPT-4o-mini interpreter described in Section~\ref{sec:physio-interpreter}, with no additional prompt or parameter tuning during evaluation, treated as a practical, instruction-following regressor; best-effort outputs for all windows.

Performance is measured using:
\begin{itemize}[leftmargin=1em, labelsep=0.3em, itemsep=2pt]
  \item \textit{MAE} for HR (BPM) and SpO\textsubscript{2} (\% points) against dataset references.
  \item \textit{Activity accuracy} from the confusion matrix.
  \item \textit{Data availability} for the conventional HR/SpO\textsubscript{2} pipeline, defined as the percentage of segments yielding valid numeric outputs. The LLM returned numeric estimates for all segments in our runs.
\end{itemize}

\subsubsection{Results and Discussion}
\label{sec:eval_results}

Table~\ref{tab:performance_metrics} summarizes results on 1003 traces. The conventional pipeline produced valid HR/SpO\textsubscript{2} outputs for 70.29\% of segments. The LLM achieved lower error and full availability: HR MAE 11.96\,BPM vs 22.49\,BPM and SpO\textsubscript{2} MAE 1.39\,\% vs 2.30\,\%. Activity accuracy was modest for both accelerometer-only methods, with a small LLM gain (38.48\% vs 32.80\%).

\begin{table}[t]
  \centering
  \caption{Performance metrics (N = 1003 traces).
           \textsuperscript{*}MAE for the conventional algorithm computed on segments with valid outputs.}
  \label{tab:performance_metrics}
  \resizebox{\columnwidth}{!}{%
    \begin{tabular}{@{}p{0.48\linewidth} >{\centering\arraybackslash}p{0.32\linewidth} >{\centering\arraybackslash}p{0.29\linewidth}@{}}
      \toprule
      \textbf{Metric} & \textbf{Conventional} & \textbf{LLM (GPT-4o mini)} \\
      \midrule
      Heart Rate MAE (BPM) & 22.49\textsuperscript{*} & 11.96 \\
      SpO\textsubscript{2} MAE (\%) & 2.30\textsuperscript{*} & 1.39 \\
      HR/SpO\textsubscript{2} Data Availability (\%) & 70.29 & 100.00 \\
      Activity Classification Accuracy (\%) & 32.80 & 38.48 \\
      \bottomrule
    \end{tabular}%
  }
\end{table}

\begin{figure*}[t]
  \centering
  \includegraphics[width=1\textwidth]{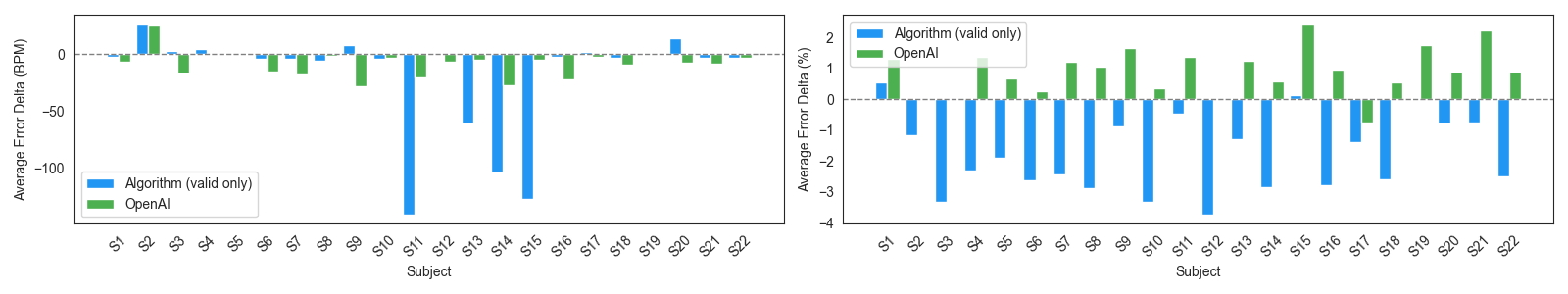}
  \caption{Average error delta for heart rate (BPM, left) and SpO\textsubscript{2} (\%, right) by subject (S1 - S22) in the Pulse Transmit Time Dataset. \textcolor{green!70!black}{LLM (green)}, and conventional estimates (\textcolor{blue}{blue}).}
  \label{fig:hr_timeseries}
\end{figure*}

\begin{figure}[htbp]
  \centering
  \includegraphics[width=\columnwidth]{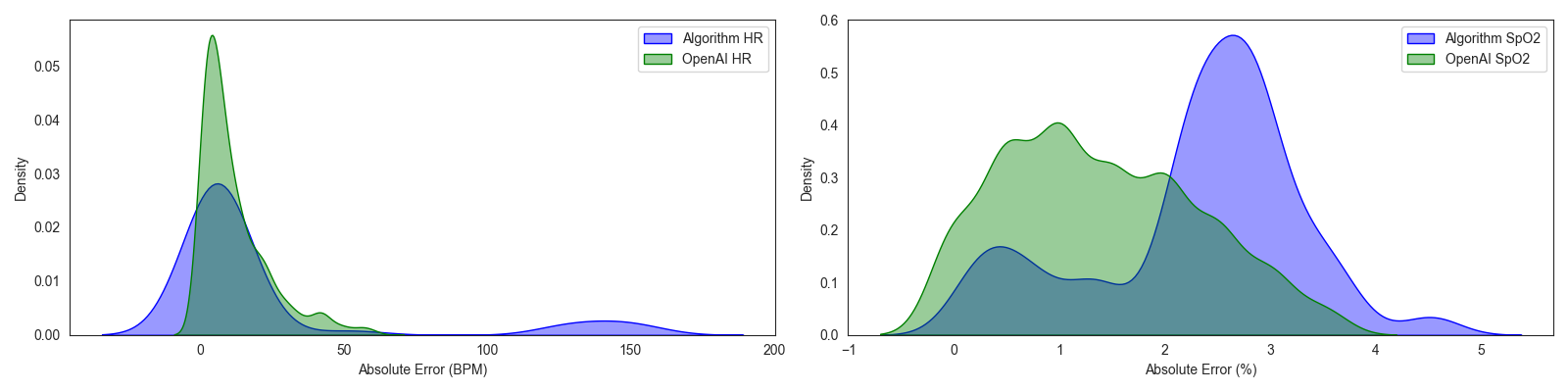}
  \caption{Density plots of absolute errors for PPG-derived metrics. Left: HR (BPM). Right: SpO\textsubscript{2} (\%). \textcolor{blue}{Conventional (blue)} vs \textcolor{green!70!black}{LLM (green)}.}
  \label{fig:error_density}
\end{figure}

\fakepar{PPG-Derived Metrics} LLM estimates for HR and SpO\textsubscript{2} are both more accurate and consistently available. Figure~\ref{fig:hr_timeseries} shows the average error by subject for both metrics; the algorithm's inability to intuitively adapt to irregular data leads to a noticeably larger error for certain subjects (S11, 13, 14, and 15) and reveals the flaw in it's deterministic design. Error densities in Figure~\ref{fig:error_density} are sharply peaked near zero for the LLM, indicating more frequent low-error predictions.

The largest gap is availability. The conventional MAX30101 pipeline, sensitive to motion and strict quality checks, emits valid HR/SpO\textsubscript{2} values for fewer segments, which limits continuous monitoring. The LLM produced estimates for all segments in this evaluation, enabling uninterrupted feedback.

\fakepar{Accelerometer-Derived Metric}
With accelerometer-only inputs, both methods achieve modest activity accuracy. The baseline LSTM shows a bias toward the \textit{run} class, often confusing \textit{walk} and \textit{sit}. The LLM yields more balanced predictions and higher recall for \textit{walk}. Future gains likely require additional motion cues such as gyroscope data or improved feature extraction.

\begin{figure}[t]
  \centering
  \includegraphics[width=\columnwidth]{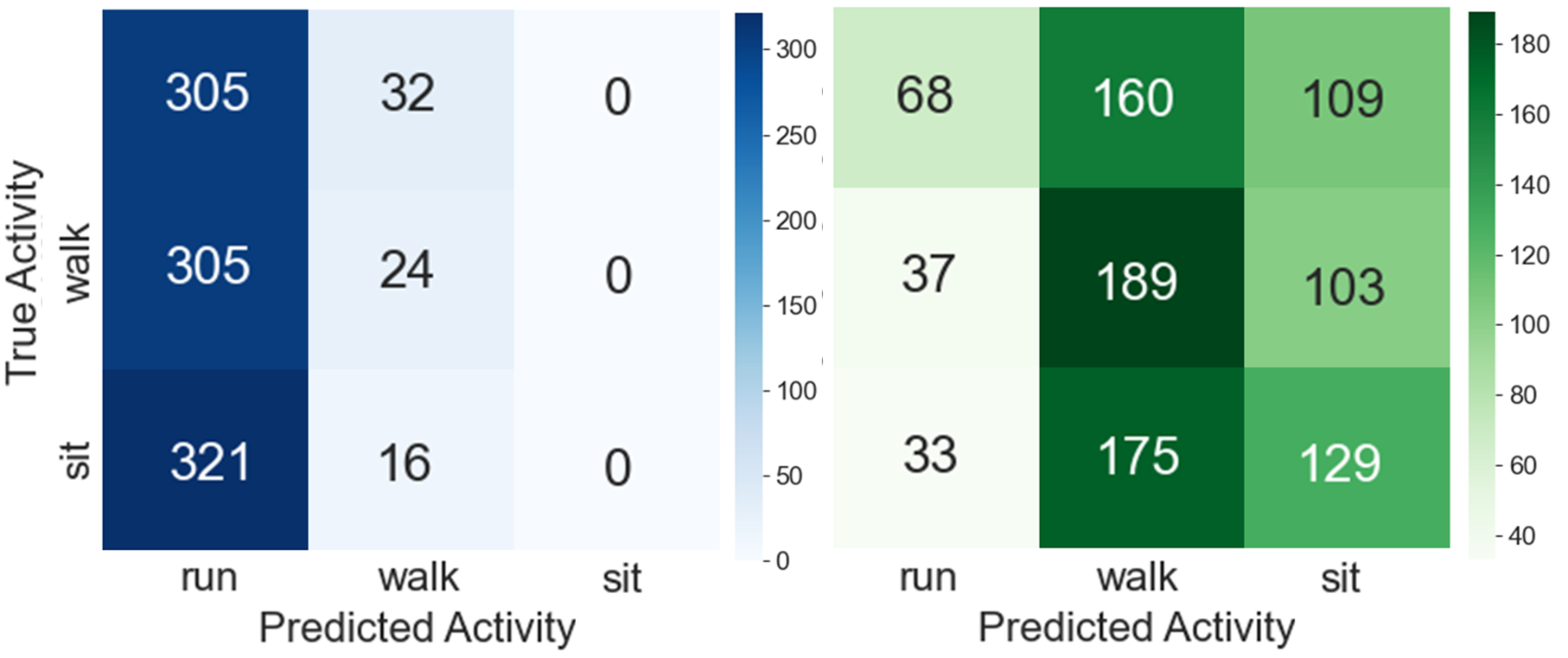}
  \caption{Confusion matrices for activity classification. Left: \textcolor{blue}{Conventional LSTM}. Right: \textcolor{green!70!black}{LLM}.}
  \label{fig:confusion_matrices}
\end{figure}


Overall, LLM delivers lower HR and SpO\textsubscript{2} error and full availability, with an advantage in activity recognition, which makes it better suited for continuous monitoring on low-cost wearables.

\subsection{Cost-Efficiency and Response Quality}
\label{sec:cost_quality_analysis}

As detailed in Section~\ref{sec:model-selection}, Guardian Angel employs an adaptive, tiered model selection strategy to optimize inference costs while maintaining response capability. We evaluate the effectiveness of this strategy using a set of 30 diverse user queries, analyzing both cost reduction and response quality preservation compared to a baseline using only the high-performance \texttt{OpenAI o1} model. 


\subsubsection{Methodology}

\fakepar{\\Cost}
To compare costs, we approximate the token count for each query using OpenAI's rule-of-thumb heuristic ($\text{Tokens} \approx \text{length}(\text{query}) / 4$)~\cite{openai_token_estimate}. The processing cost for both the tiered system (including router overhead) and the baseline was then computed based on the models used and their input token prices (Table~\ref{tab:token-pricing}), using:
\begin{equation} \label{eq:total_cost}
\footnotesize
\text{TotalCost}(q) = \sum_{m \in M_q} \left( \frac{\text{Tokens}_m(q)}{1000} \times \text{Price}_{m, \text{1K}} \right)
\end{equation}

\fakepar{Response quality}
To assess perceived quality, we generated two responses for each of the 30 queries: one from the model selected by our tiered system (\textit{Response A}) and one from the \texttt{o1} baseline (\textit{Response B}). Five independent users blindly compared these response pairs and selected their preferred one for each query, yielding 150 total comparisons. This is an exploratory, preference-only assessment (no hypothesis testing). Baseline responses were generated under a matched context.


\subsubsection{Results and Discussion}

The adaptive model selection strategy demonstrated estimated benefits in both cost and perceived quality.

\begin{figure}[t]
  \centering
  \includegraphics[width=1\linewidth]{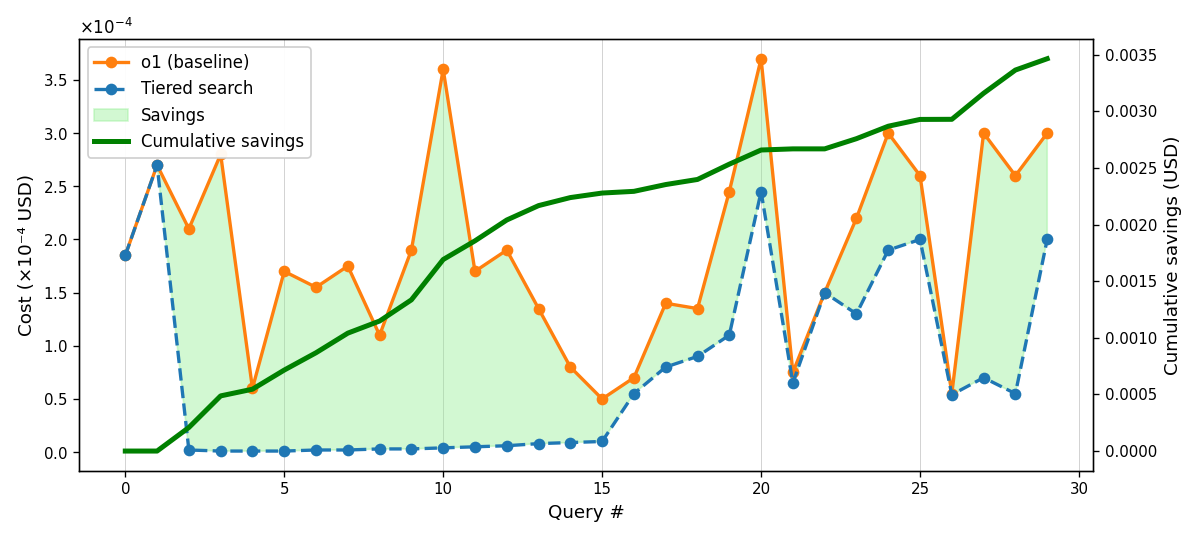}
  \caption{Cost comparison per query.}
  \label{fig:cost-lineplot}
\end{figure}

\fakepar{Cost reduction}
As summarized in Table~\ref{tab:cost-summary}, the tiered system achieved a substantial 56.57\% relative reduction in inference costs compared to the \texttt{o1}-only baseline. Figure~\ref{fig:cost-lineplot} visually confirms this, showing the tiered system's per-query cost consistently remaining below the baseline, particularly for simpler queries requiring less capable models. This ascertains the value of the tiering strategy for economic efficiency, especially in resource-constrained deployments.


\fakepar{Response quality}
Importantly, the cost savings did not come at the expense of perceived quality. In the blind user preference tests involving 150 comparisons, responses generated by the tiered model selection system were preferred 63.33\% of the time over those from the \texttt{o1}-only baseline. To ensure the comparison was fair, when the response was generated by the \texttt{o1} model, it was given the user's sensor data, as well as other relevant meta-data for context. This suggests that the tiered approach reduces costs significantly and maintains user satisfaction with the generated responses.

These results support adaptive model selection as a promising strategy, yielding significant cost savings while preserving, and often improving, the perceived quality of responses.

\begin{table}[t]
  \centering
  \scriptsize
  \begin{minipage}{0.49\linewidth}
    \centering
    \caption{Token cost).}
    \label{tab:token-pricing}
    \resizebox{0.9\columnwidth}{!}{%
    \begin{tabular}{|l|c|}
      \hline
      \textbf{Model} & \textbf{1K Tokens (USD)} \\
      \hline
      GPT\textendash4o-mini      & 0.00015 \\
      GPT-3.5-turbo              & 0.001   \\
      o3-mini-2025-01-31         & 0.00125 \\
      o1-2024-12-17              & 0.015   \\
      \hline
    \end{tabular}}
  \end{minipage}
  \hfill
  \begin{minipage}{0.49\linewidth}
    \centering
    \caption{Inference cost.}
    \label{tab:cost-summary}
    \resizebox{0.9\columnwidth}{!}{%
    \begin{tabular}{|l|c|}
      \hline
      \textbf{Metric}                & \textbf{Value (USD)} \\
      \hline
      Total cost (tiered system) & 0.0024481 \\
      Total cost (o1-only)       & 0.0056363 \\
      Total savings              & 0.0031882 \\
      Relative reduction         & 56.57\%  \\
      \hline
    \end{tabular}}
  \end{minipage}
\end{table}

\subsection{User Study}
\label{sec:user_study}
\label{sec:user-study}

To evaluate Guardian Angel’s real-world feasibility, usability, and health data comprehension, we ran a 96-hour (4-day) micro-longitudinal study.\footnote{Due to IEEE's PerCom strict policy, we could not include interview protocols, prompts, and code snippets in the appendix. We released detailed documentation through anonymized GitHub repository~\cite{guard_code}}

Phase 1 enrolled 20 university students (11 male, 9 female; age 19--29, M=21.78, SD=2.46; 50\% prior wearable experience) and produced 4,644 logged interactions. Hardware limits (five functional prototypes; Figure~\ref{fig:watches}) and IRB timelines required a tightly scoped deployment, consistent with iterative HCI practice using early discount studies to surface usability issues before larger trials~\cite{nielsen1994guerrilla, scholtz2004framework}. This convenience sample was intentionally homogeneous to isolate interface effects; generalizability is future work.

Participants were digitally skilled yet showed low health literacy typical of LMIC settings; pre-study surveys indicated uncertainty in interpreting personal physiological data~\cite{babalola2022health,lee2025challenges}. The design aimed to isolates the effect of a WhatsApp-delivered LLM on comprehension while holding digital literacy constant. Positive feasibility results motivate an IRB-approved Phase 2 with 60 participants across three age and literacy strata, following a staged deployment model suited to LMIC contexts~\cite{perrins2024health}.

\subsubsection{Study Protocol and Data Collection}
\label{sec:eval_methodology_user}
\begin{figure}[t]
  \centering
  \includegraphics[width=\columnwidth]{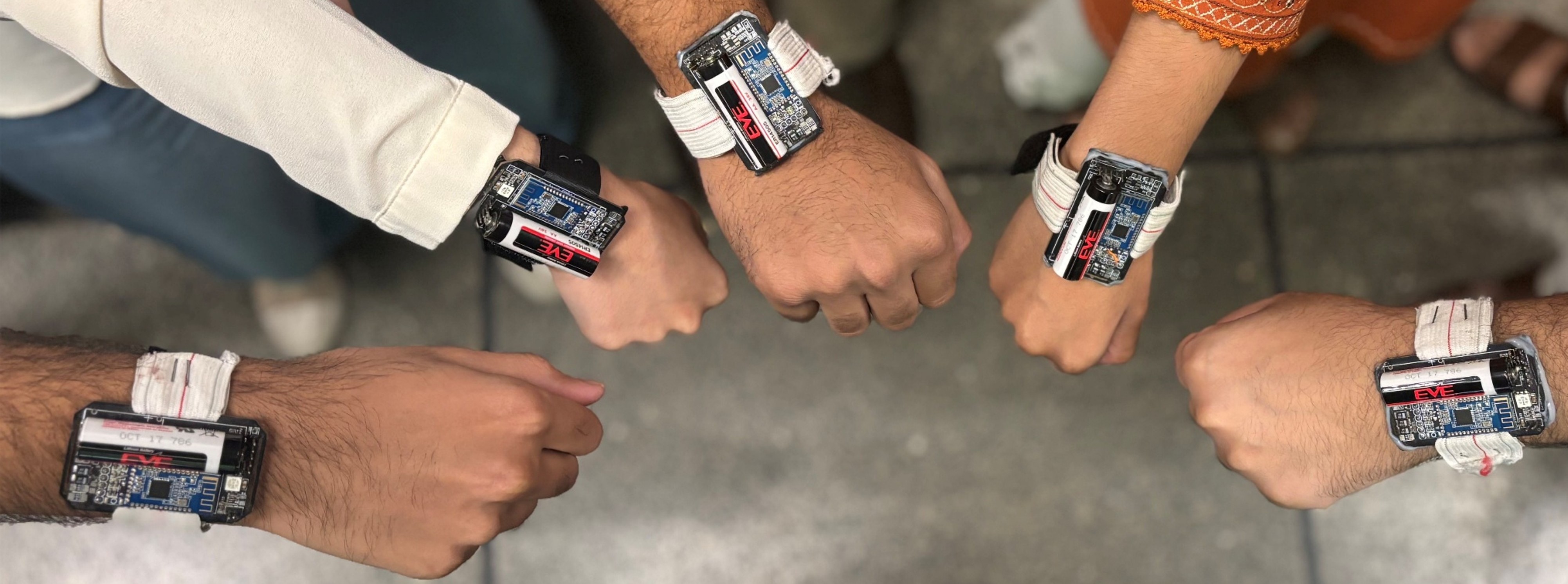}
  \caption{Five replicas of the wearable device used in the user study.}
  \label{fig:watches}
\end{figure}

Participants completed a baseline survey and semi-structured interview covering demographics, technology habits, attitudes toward AI and health tracking, and self-rated understanding of health metrics. We built five identical devices (Figure~\ref{fig:watches}) and deployed in sequential batches of five participants. During the 96-hour study, participants wore a device and interacted with the LLM over WhatsApp. System logs captured interaction frequency, timing, query type, and wear time. Post-deployment, participants completed a follow-up survey and interview on attitudes, understanding, self-reported behaviors, usability, and perceived impact. All procedures used informed consent and anonymization. Pre/post Likert responses (5-point) were analyzed as paired data; interviews underwent thematic analysis.

\subsubsection{System Engagement and Usage}
\begin{figure*}[t]
  \centering
  \includegraphics[width=\textwidth]{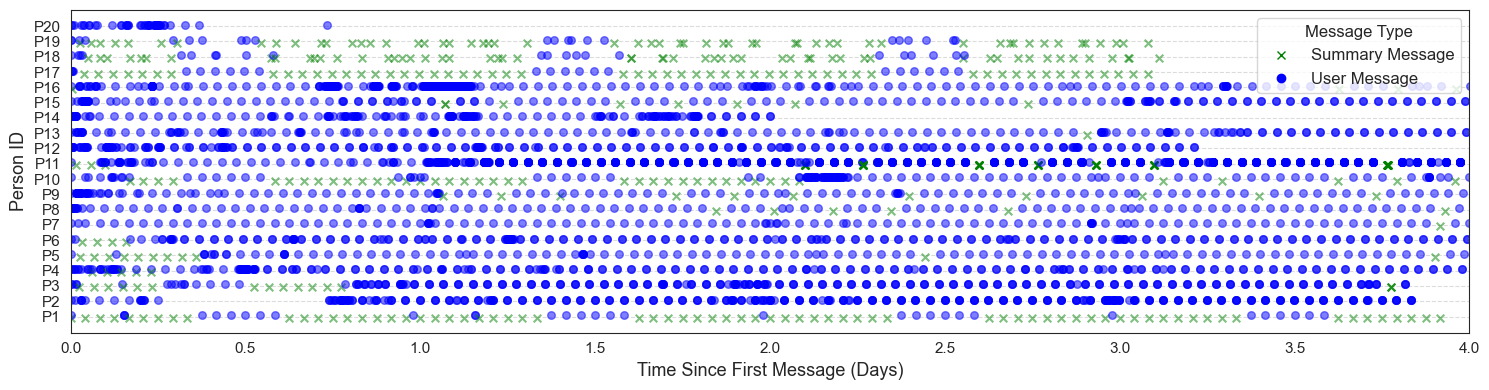}
  \caption{User messages (\textcolor{blue}{blue}) and automated summaries (\textcolor{green!70!black}{green}) for participants P1--P20 over the 96-hour study.}
  \label{fig:user-timeline}
\end{figure*}

Interaction logs indicate strong engagement. Follow-up questions were suggested on average 22 times per participant, with 12\% uptake. Users interacted in bursts (mean 9.1 sessions/day, SD=2.7). Figure~\ref{fig:user-timeline} shows user-initiated messages (blue) and automated summaries (green). Some participants experienced peripheral issues that reduced message flow, including a faulty Bluetooth link (P20) and a battery failure (P14). Others scheduled additional summaries (P1, P17, P18, P19) or turned summaries off in favor of ad-hoc queries (P7, P11, P12). Wear adherence averaged 7.3 hours/day (SD=1.8), mainly during waking hours.

Users most often requested HR and SpO\textsubscript{2}, temperature, and activity information, along with on-demand graphs and guidance from the agent. Table~\ref{tab:interaction_summary} summarizes user and system activity across all 20 participants.

\begin{table}[t]
\centering

\caption{Summary of interactions during the 96-hour (N=20).}
\label{tab:interaction_summary}
\resizebox{\columnwidth}{!}{%
\begin{tabular}{|l|l|r|}
\hline
\textbf{Category} & \textbf{Interaction Type / Detail} & \textbf{Count / Value} \\
\hline
\textbf{User-Initiated} & Voice Messages Sent & 84 \\
\cline{2-3}
 & Text Messages Sent & 554 \\
\hline
\textbf{System-Generated} & Scheduled Summaries Delivered & 2,030 \\
\cline{2-3}
 & Recommendations / Suggestions Given & 1,976 \\
\hline
\textbf{Overall Metric} & Total Logged Interactions & 4,644 \\
\hline
\end{tabular}}
\end{table}

\subsubsection{Quantitative Survey Results}
Pre/post surveys show positive shifts (Table~\ref{tab:survey_combined}). The largest gains were in accessibility of health data (+1.85), and mindfulness of HR (+1.70) and SpO\textsubscript{2} (+1.60). Additional improvements include trust in AI (+0.55), importance of personalized insights (+0.85), mindfulness of movement (+0.85), and health data comprehension (+0.55). Results suggest short-term use increased awareness, perceived value, and the ability to interpret personal health information through the conversational interface.

\begin{table}[t]
  \centering
  \caption{Pre- vs.\ post-deployment survey results.}
  \label{tab:survey_combined}
  \resizebox{\columnwidth}{!}{%
    \begin{tabular}{|p{0.58\linewidth}| >{\centering\arraybackslash}p{0.22\linewidth}|
                                    >{\centering\arraybackslash}p{0.22\linewidth}|
                                    >{\centering\arraybackslash}p{0.10\linewidth}|}
      \hline
      \textbf{Measure}
        & \textbf{Pre (M $\pm$ SD)}
        & \textbf{Post (M $\pm$ SD)}
        & \textbf{$\Delta$} \\ \hline

      \multicolumn{4}{|l|}{\textbf{Device-Related Attitudes}} \\ \hline
      Trust in AI                            & 3.25 $\pm$ 0.91 & 3.80 $\pm$ 0.77 & +0.55 \\ 
      Perceived Usability                    & 4.15 $\pm$ 1.09 & 4.30 $\pm$ 0.73 & +0.15 \\
      Importance of Personalized Insights    & 3.00 $\pm$ 1.12 & 3.85 $\pm$ 0.88 & +0.85 \\
      Accessibility of Health Data           & 2.15 $\pm$ 1.14 & 4.00 $\pm$ 0.65 & +1.85 \\ \hline

      \multicolumn{4}{|l|}{\textbf{Personal Changes}} \\ \hline
      Proactivity (Self-Reported Behavior Change) & 2.90 $\pm$ 1.37 & 3.00 $\pm$ 1.38 & +0.10 \\
      Mindfulness of Daily Movement \& Activity   & 3.10 $\pm$ 1.52 & 3.95 $\pm$ 0.95 & +0.85 \\
      Mindfulness of Heart Rate (HR)              & 2.15 $\pm$ 1.23 & 3.85 $\pm$ 0.81 & +1.70 \\
      Mindfulness of Blood Oxygen (SpO\textsubscript{2}) & 1.80 $\pm$ 1.15 & 3.40 $\pm$ 1.10 & +1.60 \\
      Mindfulness of Body Temperature             & 2.25 $\pm$ 1.16 & 3.45 $\pm$ 1.05 & +1.20 \\
      Self-Rated Physical Activity Level          & 2.85 $\pm$ 0.99 & 3.05 $\pm$ 0.95 & +0.20 \\
      Health Data Comprehension                   & 3.05 $\pm$ 1.00 & 3.60 $\pm$ 1.19 & +0.55 \\ \hline
    \end{tabular}%
  }
\end{table}

\subsubsection{Qualitative Insights from Interviews}
Thematic analysis adds context:
\begin{itemize}[leftmargin=1em, labelsep=0.3em, itemsep=2pt, topsep=0pt]
  \item \textbf{Awareness and action.} Participants reported greater awareness of physiological states and valued concise, actionable guidance (for example, hydrate, move, rest).
  \item \textbf{Conversational usability.} WhatsApp reduced learning burden and supported natural language, voice input, and code-switching; brief prompt acclimation was common and short-lived.
  \item \textbf{Feature value.} On-demand graphs and personalized reminders were frequently cited as useful and engaging.
\end{itemize}

\subsubsection{Design Principles for Resource-Constrained Wellness Systems}
To evaluate the feasibility, user acceptance, and behavioural impact of delivering an LLM-driven wearable via WhatsApp, we conducted post-study interviews. Grounded in our 1920 participant-hour field study, we distill three design principles that generalize beyond Guardian Angel:

\fakepar{(P1) Ease of Use}
Participants praised WhatsApp’s familiarity: it required no new app learning, integrated seamlessly with existing notifications, and accepted mixed-language input.  

\begin{chatlogexcerpt}[green][Participant 1] \chatline{That was very, that was easy to use because WhatsApp is something we use a lot and its notifications also come through...}{P1}
\end{chatlogexcerpt}

Fifteen respondents echoed this view. They liked rich, daily summaries yet preferred concise replies to on-demand questions:  
\begin{chatlogexcerpt}[green][Participant 1] \chatline{[Summaries had a] detailed breakdown … bullet points really helped … readability was very easy.}{P1}
\end{chatlogexcerpt}

Flexible language support further smoothed interaction:  

\begin{chatlogexcerpt}[green][Participant 3] \chatline{It captured what I'm asking well even if I asked in my native language or English.}{P3}
\end{chatlogexcerpt}

\fakepar{(P2) Perceived Value} 
Users valued actionable nudges over lengthy coaching, but wanted accuracy to sustain trust.  

\begin{chatlogexcerpt}[green][Participant 7] \chatline{The one-liner that was short-term … like drink water, sit down … was fine.}{P7}
\end{chatlogexcerpt}

\begin{chatlogexcerpt}[green][Participant 18] \chatline{It told me to hold my breath and do arm movements … I now do these light exercises daily.}{P18}
\end{chatlogexcerpt}

A few preferred broad guidance that encouraged reflection instead of prescriptive detail:  

\begin{chatlogexcerpt}[green][Participant 13] \chatline{Generalized advice leaves room for interpretation … people should do a little bit more on their own.}{P13}
\end{chatlogexcerpt}

\fakepar{(P3) Health Awareness}  
Before the study, many relied on “gut feeling.” Continuous data often contradicted those instincts and prompted reflection.  

\begin{chatlogexcerpt}[green][Participant 9] \chatline{I always go with my gut feeling … the wearable helps me infer what I need to work on.}{P9}
\end{chatlogexcerpt}

\begin{chatlogexcerpt}[green][Participant 13] \chatline{My judgment is quite wrong. When the sensors tell me my heart rate is high, I'm shocked.}{P9}
\end{chatlogexcerpt}

Data helped users connect metrics to concrete actions and spot sedentary routines:  

\begin{chatlogexcerpt}[green][Participant 4] \chatline{Knowing my heart rate helps me link it with blood glucose … then take steps for it.}{P4}
\end{chatlogexcerpt}

\begin{chatlogexcerpt}[green][Participant 5] \chatline{I realised I’m sedentary on weekends … this nudged me to move more.}{P5}
\end{chatlogexcerpt}

\fakepar{Outlook}  
Most participants expressed willingness to keep using the system if sensors and feedback were refined:

\begin{chatlogexcerpt}[green][Participant 2] \chatline{I liked it … it felt casual. If this WhatsApp thing were developed a bit more, then yes, I wouldn't mind it.}{P2}
\end{chatlogexcerpt}

Together, these insights show that a WhatsApp-based, LLM-mediated wearable can lower engagement barriers, build situational trust, and foster healthier habits with minimal user effort.

\section{Related Work and Discussion} 
\label{sec:related_work}
The convergence of mobile devices, wearables, and AI has catalysed digital health for physiology monitoring, behaviour support, and tailored interventions. We focus on two strands most relevant to ours: digital behaviour-change interventions and passive sensing via mobile or wearable devices.

\fakepar{Digital interventions for health behaviour change}
Digital behaviour-change interventions (DBCIs) scale widely and cost-effectively, yet sustaining engagement remains difficult \cite{Lee2024}. Predictors such as self-efficacy explain only part of participation variance \cite{Lee2024}. Chatbots add interactivity and have improved diet in young adults \cite{Ashton2023} and physical activity in older adults \cite{wiratunga2020fitchatconversationalartificialintelligence}. Design factors such as dialogue style, barrier reduction, and social support shape acceptance; co-creation is especially important for older adults \cite{Ashton2023, wiratunga2020fitchatconversationalartificialintelligence}. Voice interfaces can motivate use but need careful conversational design \cite{wiratunga2020fitchatconversationalartificialintelligence}. Evidence is promising \cite{Arakawa2024,Yang2024}, although superiority over simpler modalities for long-term change is unproven. Delivery mode matters: a high-engagement SMS programme for community health workers in Vietnam was feasible but did not improve knowledge, indicating that information push alone is insufficient \cite{article}.

\fakepar{Mobile and wearable sensing for health monitoring}
Passive sensing offers continuous, low-burden tracking outside clinics. For example, fusing smartphone usage with wearable signals improves sleep detection in the wild~\cite{martinez2020improved}, and privacy-aware mHealth platforms are emerging for broader populations \cite{Fernandes2024}. Key challenges include throughput, user burden, and analytics to fuse heterogeneous streams \cite{10.1145/3675094.3677583, Fernandes2024}. Recent PerCom evidence shows data interruptions and charging behavior materially impact real-world datasets” \cite{MartinezMMNCDS20}. Non-invasive blood-pressure work illustrates the trend toward deep learning: FewShotBP adapts to individuals with one-tenth the labelled data needed for conventional fine-tuning by combining multimodal features with physiological priors \cite{10.1145/3610918}.

Overall, gaps persist in sustained engagement \cite{Lee2024,Ashton2023}, theory-grounded interventions \cite{Ashton2023,wiratunga2020fitchatconversationalartificialintelligence}, effects beyond information delivery \cite{article}, and personalised models from limited real-world data \cite{10.1145/3610918,10.1145/3675094.3677583}. Guardian Angel addresses these gaps by coupling ultra-low-cost, screenless wearables with an LLM-powered WhatsApp agent, targeting affordability, digital-literacy barriers, interface complexity, and data comprehension in low-resource settings. The contribution lies in the integration, not isolated components.

\fakepar{Outlook} 
Our chosen LLM covers general health topics but lacks domain depth and adds serving cost with reliance on a third-party provider. Future work will evaluate lower-cost, open-source and sensor-language foundation models \cite{zhang2025sensorlmlearninglanguagewearable}, edge- or on-device–optimized variants, and domain-specific options (e.g., Med-PaLM) when accessible and economical for this use case. Lightweight personalization, such as small on-device adapters or profiles, is a priority to tailor guidance without increasing cost.

The hardware prioritizes low cost but leaves room for refinement. Future iterations may add sensors such as a gyroscope for richer activity features or electrodermal activity for stress, with careful budgeting for cost and battery life. \textit{Guardian Angel is a personal wellness aid, not a medical instrument}; its values are not clinical-grade and must not be used for diagnosis or treatment. The primary aim is awareness through trends and conversational feedback. Based on user feedback, we will iterate on materials and attachment methods for comfortable day–night wear and improved durability (\textit{e.g.}, dust and water resistance).


\section{Conclusion}
Guardian Angel demonstrates that pairing screenless sensing with a familiar chat interface can meaningfully extend wearable health technology in LMICs. By co-designing a simplified edge with an adaptive, WhatsApp-based LLM backend, it tackles three entrenched barriers: device cost, limited digital literacy, and the cognitive load of raw metrics. We intentionally favor a continuous physiological narrative over clinical-grade precision, where conventional pipelines drop noisy segments, Guardian Angel sustains uninterrupted, comprehensible feedback, turning failure points into engagement. Our exploratory field deployment showed sustained use and clearer understanding of health metrics, suggesting a practical path from inexpensive hardware to a trusted wellness aid.





\bibliographystyle{IEEEtran}
\bibliography{IEEEabrv, references}








\end{document}